\documentclass[%
 aip,
 amsmath,amssymb,
 reprint,%
]{revtex4-1}

\usepackage{graphicx}
\usepackage{dcolumn}
\usepackage{bm}
\usepackage{booktabs}
\usepackage{siunitx}

\usepackage[utf8]{inputenc}
\usepackage[T1]{fontenc}
\usepackage{mathptmx}
\usepackage{etoolbox}
\usepackage[english,french]{babel}

\makeatletter
\def\@email#1#2{%
 \endgroup
 \patchcmd{\titleblock@produce}
  {\frontmatter@RRAPformat}
  {\frontmatter@RRAPformat{\produce@RRAP{*#1\href{mailto:#2}{#2}}}\frontmatter@RRAPformat}
  {}{}
}%
\makeatother

\begin{document}

\preprint{AIP/123-QED}

\title{Early onset of half-power scaling laws in turbulent thermomagnetic convection}

\author{Paolo Capobianchi}
\affiliation{James Weir Fluid Laboratory, Mechanical and Aerospace Engineering, University of Strathclyde.}


\begin{abstract}
Scaling laws for turbulent thermomagnetic convection are derived through a theoretical approach supported by direct numerical simulations in a two-dimensional square cavity filled with a high-$\mathrm{Pr}$ magnetic fluid. A transport regime consistent with half-power scalings, $\mathrm{Nu} \sim \mathrm{Ra}_m^{1/2}$ and $\mathrm{Re} \sim \mathrm{Ra}_m^{1/2}$, emerges at the early onset of the laminar-to-turbulent transition and persists over the entire range investigated. Energy-budget and thermo-fluid analyses indicate that this behavior originates from the coupling between thermal and magnetic gradients, which enhances magnetic-force-driven plume ejection and vertical advection across the fluid bulk.
\end{abstract}

\maketitle

Magnetic fluids exhibit unique properties that lead to phenomena otherwise negligible in conventional fluids, such as the well-known Rosensweig instability~\cite{Cowley_Rosensweig_1967}. 
In the presence of thermal gradients and magnetic fields, the temperature dependence of the magnetization produces a non-uniform magnetic body force. 
While a purely hydrostatic state can exist, beyond a critical threshold, thermal perturbations trigger convective currents analogous to the classical Rayleigh--B\'enard (RB) instability~\cite{benard1901,Rayleigh1916}. 
In this regime, known as thermomagnetic convection (TMC)~\cite{Finlayson1970}, magnetic buoyancy replaces gravity as the primary driving mechanism. 
TMC has received significant attention in recent years owing to its ability to actively manipulate thermal fields, thereby offering a viable route for enhanced heat transfer applications (see, e.g.,~\cite{Yamaguchi1999,KRAKOV2002209,YAMAGUCHI2002203,GANGULY200463,Couto2007,Engler_2008,SZABO2018116,ALEGRETTI2024107846} and references therein). 
Despite the growing scientific interest in magnetic fluids observed in the last two decades, many fundamental aspects of TMC remain poorly understood. 
By contrast, there is a vast body of literature on turbulent natural convection spanning over a century, starting from seminal formulations~\cite{Oberbeck1879,Boussinesq1901}, evolving through foundational scaling theories~\cite{Malkus1954,Priestley1954,chandrasekhar2013hydrodynamic,kraichnan1962,spiegel1971,GROSSMANN_LOHSE_2000}, extensive experiments~\cite{Castaing_Gunaratne_Heslot_Kadanoff_Libchaber_Thomae_Wu_Zaleski_Zanetti_1989,Chavanne1997,Couto2007}, and recent high-resolution Direct Numerical Simulations (DNS)~\cite{Johnston2009,SHISHKINA_WAGNER_2006,van_der_Poel_Stevens_Lohse_2013,Zhu2018}; while many aspects of turbulent RB convection are reasonably well understood nowadays, the ultimate regime predicted theoretically by~\citet{kraichnan1962} has left open questions, despite the massive efforts dedicated to the subject~\cite{Lohse_Shishkina1}. A comprehensive account of the fundamental RB convection literature can be found in~\cite{Ahlers_et_al_2009}.

Unlike classical RB convection, turbulent thermomagnetic convection has received remarkably little attention, and to our knowledge, no prior studies have addressed purely magnetically driven turbulent regimes. Arguably, upon the transition to turbulence, the convective thermal efficiency, measured by the Nusselt number $\mathrm{Nu}$, is anticipated to significantly exceed that of the laminar case. To address this, we carried out a series of DNS of a turbulent TMC flow in a square cavity filled with a ferrofluid with a Prandtl number $\Pr\approx 50$, which allowed us to determine scalings for the Nusselt and Reynolds ($\mathrm{Re}$) numbers as a function of the magnetic Rayleigh number, $\mathrm{Ra}_m$. While acknowledging the limitations of two-dimensional turbulent approximations, we adopt a 2D domain to mitigate computational costs, building upon the well-documented analogy with turbulent RB systems~\cite{van_der_Poel_Stevens_Lohse_2013}. In tandem with the computations, we derived these scaling laws through a general theoretical framework inspired by the Grossmann--Lohse theory~\cite{GROSSMANN_LOHSE_2000} originally formulated for the classical Rayleigh--Bénard problem, suitably modified for thermomagnetic convection.

Overall, we assessed the performance of the system across a range of magnetic Rayleigh numbers covering nearly two orders of magnitude immediately after the onset of turbulence, as further explained below. Remarkably, we show that immediately following the turbulent onset, the system evolves into a state where both $\mathrm{Nu}$ and $\mathrm{Re}$ scale with an exponent approximately $1/2$. This behavior is in good agreement with our general theoretical analysis, notwithstanding that the latter relies on a classical three-dimensional framework. This agreement suggests that the two-dimensional approximation captures the dominant transport trends, as far as mean transport quantities are concerned. Intriguingly, this scaling resembles the theoretical predictions of~\citet{kraichnan1962}, in its dependence on the Rayleigh number, irrespective of the Prandtl-number dependence, although it emerges while the boundary layers are still laminar, suggesting that  magnetic forcing can bypass the diffusive limitations inherent to Malkus's theory for standard RB convection~\cite{Malkus1954}.

With the above in mind, we consider the problem of a magnetic fluid enclosed in a square cavity of side $L$. A temperature difference $\Delta \theta = \beta L$ is imposed by prescribing the temperature, $\theta$, at two opposite walls, while a uniform magnetic field of strength $H_0$ aligned with the temperature gradient is maintained; gravity is neglected throughout. The magnetic field is established by prescribing different magnetic potentials, $\phi_m$, at the heated boundaries, ensuring infinite magnetic permeability at the walls, matching the boundary conditions used in the linear stability analysis of \citet{Gotoh-Yamada_1982}. The lateral walls are adiabatic and magnetically insulated. Following~\cite{Finlayson1970}, the magnetization, $M_i= M \left( H,\theta\right) e_{Hi}$, is linearized about $H_0$ and the average temperature $\theta_0$:
\begin{equation}
\label{eqn:linM}
    M\left(H,\theta \right)=M_0+\chi\left( H-H_0\right)-K\left(\theta-\theta_0 \right),
\end{equation}
where $M_0$ is the equilibrium magnetization at the temperature $\theta_0$; $\chi$ and $K$ are the magnetic susceptibility and pyromagnetic coefficient, respectively; and $e_{Hi}=H_i/H$. 
The magnetic field is governed by the magnetostatic Maxwell equations (see, e.g.,~\cite{Capobianchi}):
\begin{equation}
\epsilon_{ijk} \partial_j H_k = 0, \quad \partial_i B_i = 0
\label{eqn:Maxwell_indicial},
\end{equation}
where $B_i = \mu_v(H_i + M_i)$ is the magnetic induction and $\mu_v$ the vacuum permeability. Thermal gradients within the flow generate a Kelvin body force, $\mu_v M \partial_i H$~\cite{neuringer1964ferrohydrodynamics}, which can be redefined via Eq.~\eqref{eqn:linM} to highlight the magnetic pressure contribution, $p_H=\mu_v M_0H+\frac{1}{2}\mu_v\chi\left(H-H_0\right)^2$ (see, e.g.,~\cite{suslov2008}) such that:
\begin{equation}
\begin{aligned}
\mu_v M \partial_i H &= \partial_i p_{H} - \mu_v K\left(\theta-\theta_0 \right) \partial_i H. 
\end{aligned}
\label{eqn:pressure}
\end{equation}

The flow is governed by the incompressibility constraint, $\partial_i u_i = 0$, along with the momentum, energy, and magnetic field equations~\cite{suslov2008}. These are nondimensionalized using the cell height $L$, the thermal diffusivity $\kappa$, the temperature difference, $\theta-\theta_0$, and the reference magnetic field $K\beta L/(1+\chi)$:
\begin{gather}
\partial_t \hat{u}_i + \hat{u}_j \partial_j \hat{u}_i = -\partial_i \hat{P} + \mathrm{Pr} \partial_{jj} \hat{u}_i - \mathrm{Pr} \, \mathrm{Ra}_m \hat{\theta} \partial_i \hat{H}, \label{eqn:momentum} \\ \partial_t \hat{\theta} + \hat{u}_i \partial_i \hat{\theta} = \partial_{ii} \hat{\theta}, \label{eqn:energy} \\ \partial_i \hat{H}_i + \hat{e}_{Hi} \partial_i \hat{\theta} = 0, \label{eqn:Maxwell} 
\end{gather}
where $(\hat{\cdot})$ denotes dimensionless variables, $\hat{P} = \hat{p} - \hat{p}_H$ is the modified pressure and $\hat{\theta} = \mp 1/2$ at the heated boundaries ($\hat{z} = \pm 1/2$). The magnetic Rayleigh number, defined as $\mathrm{Ra}_m = \mu_v \beta^2 K^2 L^4 / [\rho \nu \kappa (1+\chi)]$, where $\rho$ and $\nu$ are density and kinematic viscosity, quantifies the relative importance of thermomagnetic buoyancy compared to molecular diffusion; $\mathrm{Pr}=\nu/\kappa$. According to the notation used in~\cite{Gotoh-Yamada_1982}, the nonlinearity of magnetization can be accounted for by the parameter $M_3 = B_0 / [\mu_v H_0 (1+\chi)]$; the present work focuses on linear magnetization conditions, in which case $M_3 = 1$. Unless otherwise stated, hereafter we consider dimensionless variables and the ``hat'' symbol will be omitted for brevity. 

The governing equations (system~\eqref{eqn:momentum}--\eqref{eqn:Maxwell} coupled with mass conservation) are solved on a fully structured grid of $400 \times 800$ cells in the $xz$-plane. The temperature gradient and magnetic field are imposed along the $z$ direction perpendicular to the $x$-axis. The grid is refined near the walls using a geometric expansion $\Delta z_i = \Delta z_w r^{i-1}$, with $i$ denoting the cell index from the wall. Here, $\Delta z_w$ is the wall-adjacent cell height and $r$ the expansion ratio. The grid anisotropy is set to $\Delta z_{mp}/\Delta z_w = 16$ at the mid-plane, ensuring that the Batchelor scale is fully resolved throughout the domain. Further details on the resolution of these scales are provided below.

Direct numerical simulations (DNS) are performed using a custom transient, pressure-based solver for incompressible flows, implemented within the OpenFOAM framework by coupling the momentum and energy equations with the Maxwell equations. Pressure-velocity coupling is handled via the PISO (Pressure Implicit with Splitting of Operators) algorithm. Validation of the solver against the linear stability analysis of \cite{Gotoh-Yamada_1982} is provided in the Supplemental Material. Convective terms are discretized using second-order accurate central difference schemes, while temporal derivatives are treated with a first-order accurate Euler scheme. The time-step is adjusted dynamically during run-time to maintain a maximum allowable Courant number, $\textrm{Co}_{\textrm{max}} = 0.1$. 
\begin{figure}[t]
\centering
\includegraphics[width=\columnwidth]{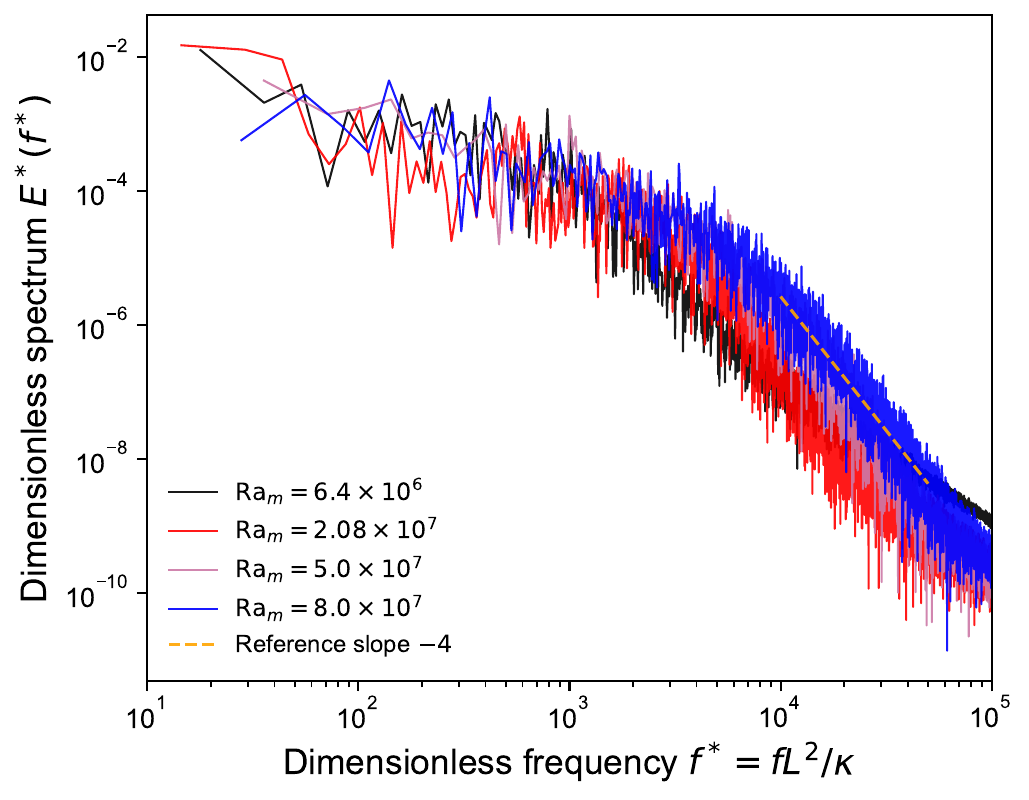}
\caption{Dimensionless turbulent kinetic energy power spectra for $\mathrm{Ra}_m=6.4\times10^6,\, 2.08\times10^7,\,4.8\times10^7,\, 8.0\times10^7$. Time and velocity are normalized using the thermal diffusion scales $L^2/\kappa$ and $\kappa/L$, respectively. Representative high-frequency power-law exponents, computed over the arbitrary interval $1.0\times10^4<f^*<5.0\times10^4$, are approximately $-2.1$, $-3.2$, $-4.6$ and $-4.3$, respectively. These values are provided only as a qualitative indication of the evolving spectral behavior and should not be interpreted as asymptotic scaling exponents. A dashed line with slope $-4$ within the range used for the fits has been added as an aid to guide the eye.}
\label{fig:Fig1}
\end{figure}
Simulations are set considering the properties of a kerosene-based ferrofluid (NF4000C, Ferrotec, USA) with a Prandtl number $\mathrm{Pr} \simeq 49.4$~\cite{HECKERT2015337}. The magnetic Rayleigh number is varied in the range $1.6 \times 10^6\leq \mathrm{Ra}_m \leq 8.0 \times 10^7 $,  corresponding to turbulent conditions (the flow was found to be laminar for $\mathrm{Ra}_m \leq 8.0\times10^5$). Each simulation is run until statistically stationary conditions are reached. Subsequently, cases are restarted and continued until the mean velocity and temperature fields are fully converged. This procedure ensured well-developed statistics for the Nusselt and Reynolds numbers, respectively calculated as $\mathrm{Nu} = -\langle \partial_z \theta \rangle_{t,A}$ and $\mathrm{Re} = u_{rms}\,L/\nu$, where $u_{rms}=\sqrt{\langle u_{i}u_{i}\rangle_{t,V}}$ is the volume-averaged root mean square velocity. Here, $\langle \cdot \rangle_{t,\,A,\,V}$ denote averages over time, heated surfaces and volume, respectively. Mesh resolution was assessed at the highest magnetic Rayleigh number, $\mathrm{Ra}_m = 8.0 \times 10^7$ by evaluating the Batchelor scale, $\lambda_B = \eta / \sqrt{\mathrm{Pr}}$, where $\eta$ is the Kolmogorov scale. The analysis confirmed that the flow is fully resolved at the boundaries, with approximately 30 cells within the thermal boundary layers, while at the mid-plane, where the grid spacing is coarsest, the cell size reaches its maximum value $\Delta s_{mp} = 16\Delta s_{w}$, yielding $\lambda_B \approx \Delta s_{mp}$. Additionally, we performed an internal consistency check on the Nusselt number by comparing the average wall flux value with that derived from the global thermal dissipation rate (see Eq.~\eqref{eqn:epsilon_theta}). The relative difference remained consistently between $1\%$ and $1.5\%$ for $\mathrm{Ra}_m=4.8 \times 10^7$ and between $2\,\%$ and $2.5\,\%$ for $\mathrm{Ra}_m=8.0\times 10^7$. Additionally, raw Nusselt-number data and the corresponding confidence intervals are provided in the Supplemental Material. The data confirm the adequacy of the spatio-temporal resolution, but also indicate that simulations at higher values of $\mathrm{Ra}_m$ would require a redesign of the mesh. Figure~\ref{fig:Fig1} shows the dimensionless turbulent kinetic energy spectra for four representative cases, revealing a systematic change in the flow dynamics with increasing $\mathrm{Ra}_m$. At high frequencies, the spectra become progressively steeper as $\mathrm{Ra}_m$ increases. The observed exponents are bounded between approximately $-2.1$ and $-4.6$, with the two largest forcing levels exhibiting the steepest slopes with similar spectral behavior. For all datasets, an energy-conservation consistency check was performed, yielding $\int E^*(f^*)\,df^* \simeq 1$ to numerical precision.

In the following, we provide the theoretical arguments, informed by the DNS data, leading to the observed scalings. First, we consider the global thermal dissipation rate, $\epsilon_{\theta}$, which is rigorously related to the Nusselt number as follows~\cite{shraiman_siggia}:
\begin{equation}
    \epsilon_{\theta} = \langle (\partial_i \theta)^2 \rangle_{V} = \mathrm{Nu}.
    \label{eqn:epsilon_theta}
\end{equation}
An expression for the kinetic dissipation rate, $\epsilon_{u}$, is obtained from the kinetic energy budget. Using the stationary Eq.~\eqref{eqn:energy}, we can determine the kinetic energy equation as:
\begin{equation}
\begin{split}
   u_j \partial_j \left(\frac{u_i u_i}{2}\right) = & - \partial_i (u_i P) + \mathrm{Pr} \, u_i \partial_{jj} u_i \\
   & - \mathrm{Pr} \, \mathrm{Ra}_m u_i \theta \partial_i H,
\end{split}
\label{eqn:kinetic}
\end{equation}
where we have taken advantage of the incompressibility constraint, $\partial_i u_i=0$. By integrating over the entire volume of the domain and accounting for the solid boundaries (no-slip and no-penetration), the convective and pressure terms vanish. After manipulating the viscous term, we obtain:
\begin{equation}
0 = - \mathrm{Pr} \langle (\partial_j u_i)^2 \rangle_V - \mathrm{Pr}\, \mathrm{Ra}_m \langle u_i\theta \partial_i H \rangle_V,
\label{eqn:energy_balance}
\end{equation}
where $\epsilon_u=\langle(\partial_j u_i)^2\rangle_V$ denotes the dimensionless viscous dissipation rate. To relate the magnetic term to the thermal field, we use the identity
\begin{equation}
u_i\theta \partial_i H = u_i \partial_i (\theta H) - H u_i \partial_i \theta.
\end{equation}
From the stationary energy equation, $u_i \partial_i \theta = \partial_{ii} \theta$, so that Eq.~\eqref{eqn:energy_balance} can be rewritten as:
\begin{equation}
 \langle (\partial_j u_i)^2 \rangle_V
 =
 - \mathrm{Ra}_m
 \left[
 \langle \partial_i (u_i\theta H) \rangle_V
 -
 \langle H \partial_{ii} \theta \rangle_V
 \right],
\label{eqn:energy_balance_II}
\end{equation}
where we have again used the incompressibility constraint. Since the domain is bounded by solid boundaries, $\langle \partial_i (u_i\theta H) \rangle_V = 0$, and therefore
\begin{equation}
\begin{split}
 \langle (\partial_j u_i)^2 \rangle_V
 &= \mathrm{Ra}_m \langle H \partial_{ii} \theta \rangle_V \\
 &= \mathrm{Ra}_m
 \left[
 \langle \partial_i (H\partial_i\theta) \rangle_V
 -
 \langle \partial_i\theta \partial_iH \rangle_V
 \right].
\end{split}
\label{eqn:energy_balance_III}
\end{equation}
In order to evaluate the first term on the right-hand side of Eq.~\eqref{eqn:energy_balance_III}, it is useful to consider first the one-dimensional conductive reference state. In this state, the magnetic field is aligned with the imposed field direction, such that
\begin{equation}
    H_i = H(z) e_{zi},
    \qquad
    e_{Hi} = e_{zi}.
\end{equation}
Equation~\eqref{eqn:Maxwell} therefore reduces to $\partial_z H+\partial_z\theta=0$, which can be integrated to give $H+\theta=H_0$, where $H_0$ is a constant. Using $\theta(\pm1/2)=\mp1/2$, we obtain $H(\pm1/2)=H_0\pm1/2$. In the convective state, however, the previous solution does not necessarily hold, owing to the fact that the temperature gradient is no longer uniform. Motivated by the structure of the diffusive solution, we parameterize the mean magnetic field difference between the horizontal boundaries as $H(-1/2)=\langle H_0\rangle_t-c/2$, and $H(1/2)=\langle H_0\rangle_t+c/2$, where $c$ is an unknown dimensionless constant characterizing the mean magnetic-field difference between the two horizontal boundaries and $\langle H_0\rangle_t$ is the time-averaged magnetic field at $z=0$; as we shall see, the background constant $\langle H_0\rangle_t$ does not affect the net surface contribution, as it cancels out when evaluating the difference between the two boundary contributions. The integral
\begin{equation}
    \left\langle \partial_i (H\partial_i\theta) \right\rangle_V
    =
    \frac{1}{V}\int_S H\partial_i\theta n_i \, dS,
    \label{eqn:surf_int}
\end{equation}
can then be partitioned over the boundaries of the domain as
\begin{equation}
\begin{split}
\int_S H\partial_i\theta n_i \, dS
={}&
\int_{S_{(-1/2)}}
\left(H_0-\frac{c}{2}\right)
\partial_i\theta n_i \, dS
\\
&+
\int_{S_{(1/2)}}
\left(H_0+\frac{c}{2}\right)
\partial_i\theta n_i \, dS,
\end{split}
\end{equation}
where the lateral boundaries do not contribute to the integral. Recalling that the dimensionless heat flux is defined as
\[
\mathrm{Nu}=-\langle\partial_z\theta\rangle_{t,A}
\]
at the walls, and considering the outward unit normals ($n_z=-1$ at $z=-1/2$ and $n_z=1$ at $z=1/2$), the surface integral yields
\begin{equation}
    \left\langle \partial_i (H\partial_i\theta) \right\rangle_V
    =-c\mathrm{Nu},
\end{equation}
for a domain of unit height. Substituting this result into Eq.~\eqref{eqn:energy_balance_III}, we obtain
\begin{equation}
 \epsilon_u
 =
 \mathrm{Ra}_m
 \left[
 -c\mathrm{Nu}
 -
 \langle \partial_i\theta \partial_iH \rangle_V
 \right].
\label{eqn:final_energy_balance}
\end{equation}
The thermal and magnetic gradients are, on average, antiparallel, so that the cross-correlation term $-\langle \partial_i\theta \partial_iH \rangle_V$ provides a positive contribution to the kinetic-energy balance. The structure of Eq.~\eqref{eqn:Maxwell} implies that magnetic and thermal gradients are intrinsically coupled, while departures from a local balance between them arise from spatial variations in the magnetic-field orientation through the term $H\partial_i e_{Hi}$. In the conductive limit, Eq.~\eqref{eqn:Maxwell} yields $-\partial_i\theta\,\partial_iH = (\partial_i\theta)^2$, suggesting that the correlation term naturally scales with the thermal dissipation rate and therefore with the Nusselt number. Motivated by this observation, we attempt a closure hypothesis by introducing the dimensionless parameter $\Lambda$ such that
\begin{equation} \Lambda \equiv \frac{- \langle \partial_i \theta \partial_i H \rangle_V}{\mathrm{Nu}} . 
\label{eqn:lambda_def}
\end{equation} 
Substituting Eq.~\eqref{eqn:lambda_def} into Eq.~\eqref{eqn:final_energy_balance} yields
\begin{equation} 
\epsilon_u = \mathrm{Ra}_m \mathrm{Nu} (\Lambda-c).
\label{eqn:epsilon_u} 
\end{equation}
The conductive reference state is recovered as a limiting case of Eq.~\eqref{eqn:epsilon_u}. In the absence of fluid motion, Eq.~\eqref{eqn:Maxwell} reduces to $\partial_z H+\partial_z\theta=0$, which implies $\Lambda=1$. At the same time, the magnetic-field difference between the horizontal boundaries is unity, i.e., $c=1$. Consequently, Eq.~\eqref{eqn:epsilon_u} yields $\epsilon_u=0$, as required for a purely conductive state. Equation~\eqref{eqn:epsilon_u} maintains a formal analogy with the classical energy balance in Rayleigh--Bénard convection~\cite{shraiman_siggia}, where viscous dissipation is likewise directly related to global heat transport. As shown below, this relation provides the basis for the scaling behavior observed in the turbulent thermomagnetic regime. DNS data suggest that $\Lambda-c$ remains of order unity and exhibits no systematic dependence on $\mathrm{Ra}_m$ within the investigated range. Consequently, it is treated as a constant in the scaling analysis that follows. It should be emphasized, however, that possible dependencies on the Prandtl number, boundary conditions, aspect ratio, and flow dimensionality (i.e., two- or three-dimensional configurations) cannot be excluded and lie beyond the scope of the present work.
\begin{table}[t]
\centering 
\caption{Average viscous ($\delta_u$) and thermal ($\delta_\theta$) boundary layer thicknesses calculated across all computational cells of both heated boundaries for all investigated $\mathrm{Ra}_m$. Thicknesses are defined as $\delta_u = 1 / |\partial_z (u/u_{\text{max}})|_w$ and $\delta_\theta = (\theta_w - \theta_b) / |\partial_z \theta|_w$, where subscript $w$ denotes wall values. The final column reports the ratio $\delta_u/\delta_\theta$.}
\label{tab:Tab1}
\hbox to \linewidth{\hss \begin{minipage}{0.75\linewidth} 
\begin{ruledtabular}
\begin{tabular}{S[table-format=1.2e1] S[table-format=1.4] S[table-format=1.4] S[table-format=1.2]}
{$\mathrm{Ra}_m$} & {$\delta_u$} & {$\delta_{\theta}$} & {$\delta_u / \delta_{\theta}$} \\
\hline
2.00e6  & 0.0386 & 0.0612 & 0.63 \\
3.20e6  & 0.0297 & 0.0344 & 0.86 \\
6.40e6  & 0.0258 & 0.0233 & 1.11 \\
1.28e7  & 0.0167 & 0.0159 & 1.05 \\
2.08e7  & 0.0150 & 0.0123 & 1.22 \\
3.20e7  & 0.0221 & 0.0100 & 2.21 \\
4.80e7  & 0.0207 & 0.0083 & 2.49 \\
8.00e7  & 0.0151 & 0.0060 & 2.52 \\
\end{tabular}
\end{ruledtabular}
\end{minipage}\hss}
\end{table}

\begin{figure*}[t]
    \centering
    \begin{minipage}[b]{0.48\textwidth}
        \centering
        \includegraphics[width=\textwidth]{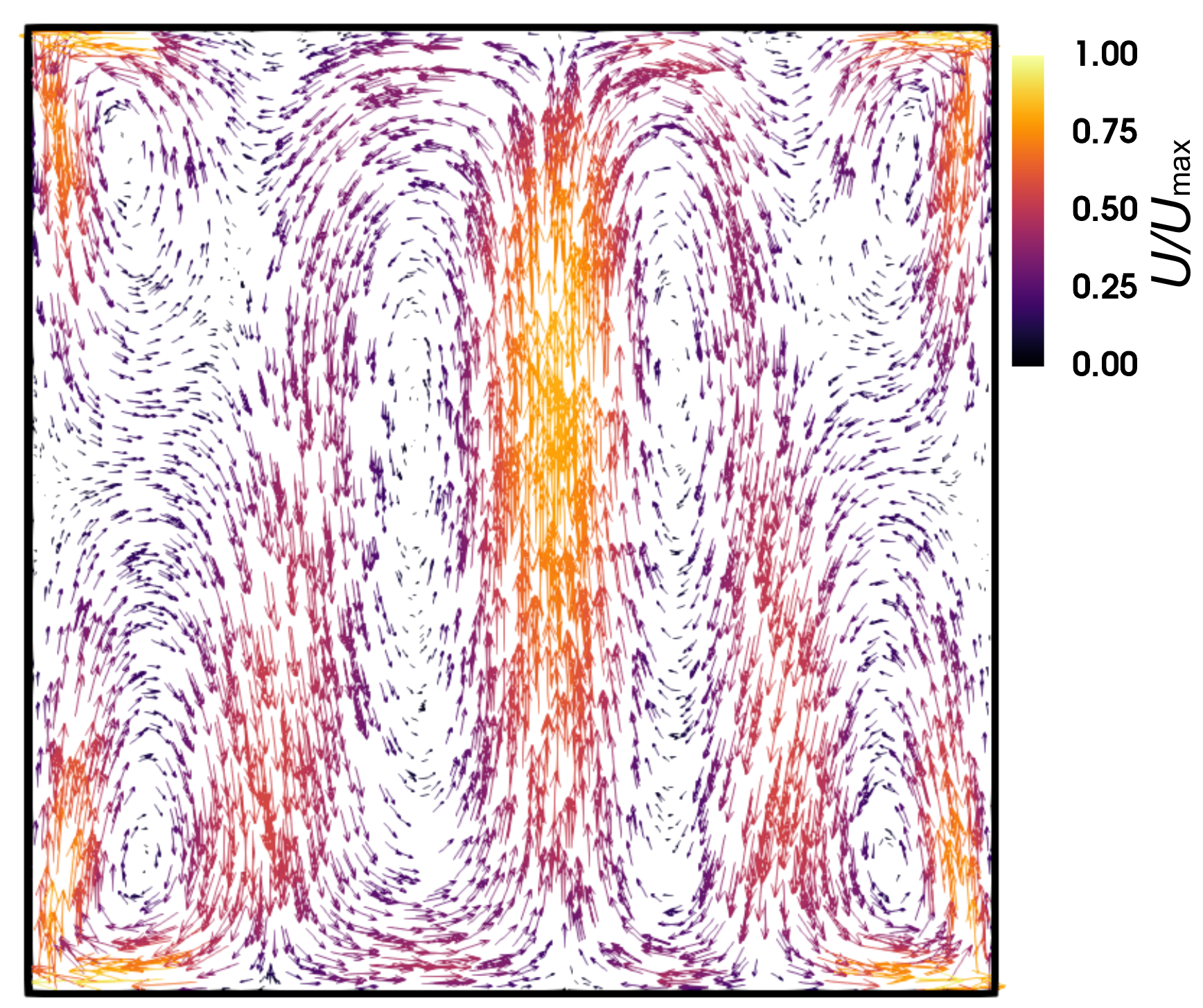} \\
        \vspace{-0.5cm} 
            \hfill \small\textsf{\textbf{(a)}} \hspace{0.6cm}
    \end{minipage}
    \hfill
    \begin{minipage}[b]{0.48\textwidth}
        \centering
        \includegraphics[width=\textwidth]{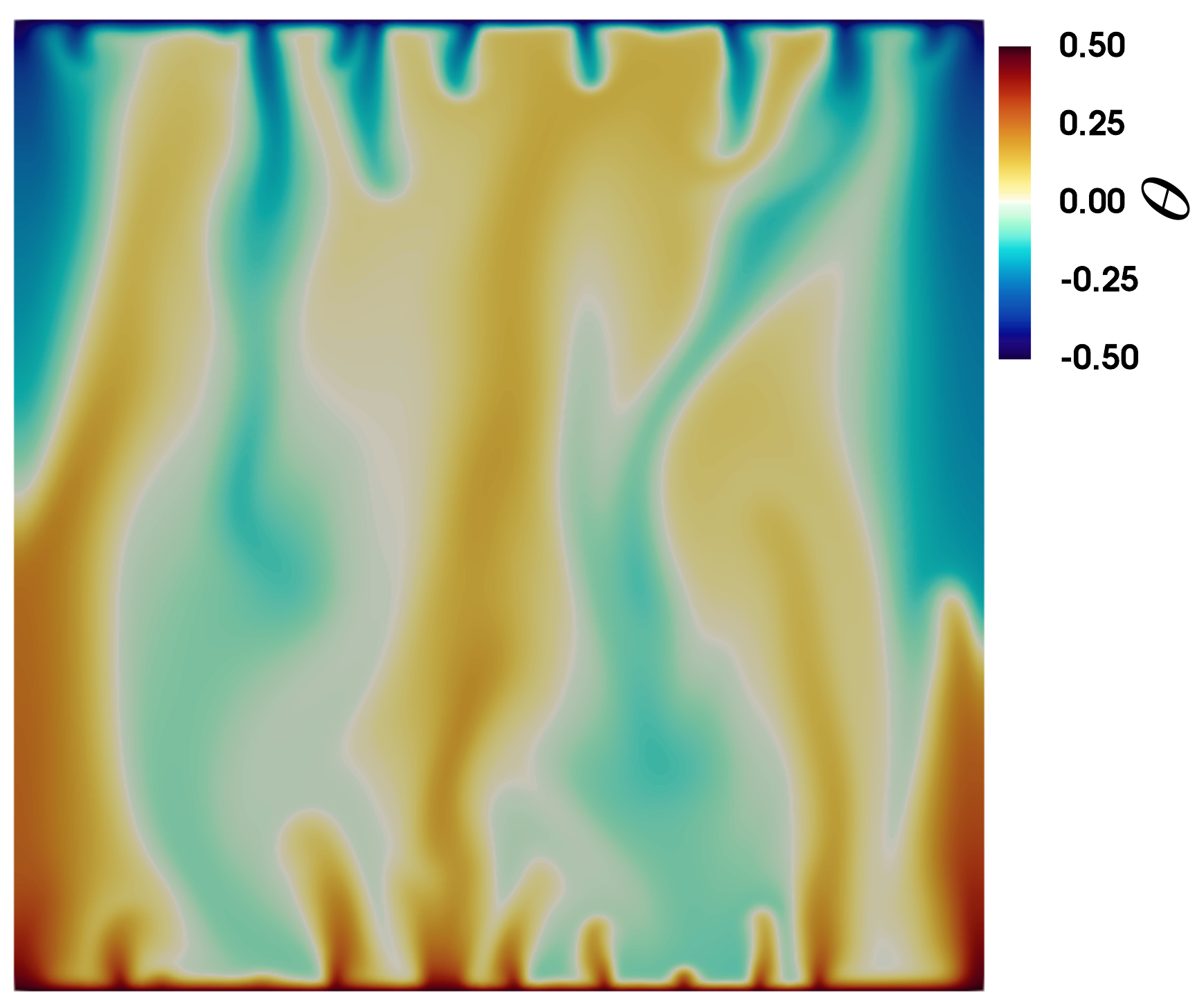} \\
        \vspace{-0.5cm}
            \hfill \small\textsf{\textbf{(b)}} \hspace{0.6cm}
    \end{minipage}
    \caption{(a) Normalized time-averaged velocity vector field $U/U_{\text{max}}$, and (b) instantaneous temperature field $\theta$. Data correspond to $\mathrm{Ra}_m=2.08\times10^{7}$.}

    \label{fig:Fig2}
\end{figure*}

\begin{figure*}[t]
    \centering
    \includegraphics[width=0.48\textwidth]{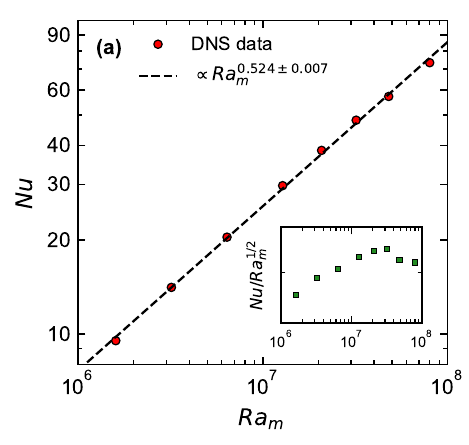}\hfill
    \includegraphics[width=0.48\textwidth]{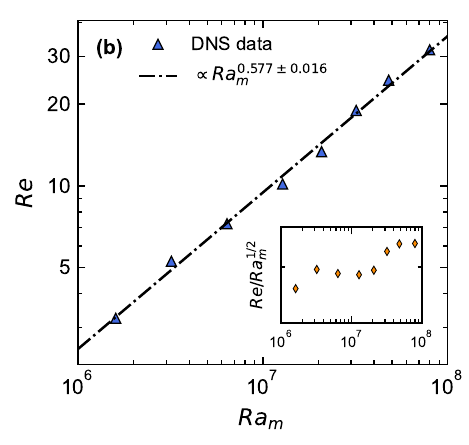}
    \caption{(a) $\mathrm{Nu}$ and (b) $\mathrm{Re}$ as functions of the magnetic Rayleigh number $\mathrm{Ra}_m$. Symbols denote the DNS data. Dashed and dash-dotted lines represent the unconstrained power-law best fits, with the optimal scaling exponents and their statistical uncertainties displayed within the legends. The insets show the corresponding compensated plots scaled by the theoretical prediction $\mathrm{Ra}_m^{1/2}$. For graphical clarity and to avoid overlap with the main panels, the vertical axes of the insets lack numerical labels; the represented intervals span from $0.0075$ to $0.0085$ for $\mathrm{Nu}/\mathrm{Ra}_m^{1/2}$ in (a), and from $0.0025$ to $0.0036$ for $\mathrm{Re}/\mathrm{Ra}_m^{1/2}$ in (b).}
    \label{fig:Fig3}
\end{figure*}

In the following, we use the approach of Grossmann and Lohse~\cite{GROSSMANN_LOHSE_2000}, determining the scalings from the viscous and thermal dissipation while treating the boundary layer and bulk contributions separately; our initial focus is on the former. Denoting by $U$ the large-scale velocity, according to~\cite{GROSSMANN_LOHSE_2000} Eq.~\eqref{eqn:epsilon_u} can be recast as:
\begin{equation}
\frac{U^2}{\delta_{u}} \sim \mathrm{Ra}_m\, \mathrm{Nu}, 
    \label{eqn:epsilon_u_delta}
\end{equation}
where $\delta_u$ is the viscous boundary layer thickness.  Table.~\ref{tab:Tab1}, which compares the viscous and thermal ($\delta_{\theta}$) boundary layer thicknesses, indicates that $\delta_u/ \delta_{\theta}\sim O(1)$, despite the large Prandtl number. Remarkably, the viscous boundary layer thinning arises from magnetic body forces acting in close proximity to the walls, effectively constraining the viscous scale to match the thermal one, as previously described by~\citet{BEDNARZ20091157}. In light of these findings, since $\mathrm{Nu} \sim 1/\delta_{\theta}$, Eq.~\eqref{eqn:epsilon_u_delta} can be recast as $U^2 \sim \mathrm{Ra}_m\,(\delta_u/\delta_\theta)$. The numerical values reported in Table~\ref{tab:Tab1} show that $\delta_u/\delta_\theta$ remains of order unity throughout the investigated range, although a gradual increase with $\mathrm{Ra}_m$ is observed. This increase appears to weaken at the largest forcing levels considered, where the ratio exhibits indications of saturation. Consequently, a weak residual dependence on $\mathrm{Ra}_m$ cannot be excluded over the lower portion of the investigated range. Such a dependence may partly contribute to the fact that the DNS exponents shown later in Fig.~\ref{fig:Fig3} remain slightly larger than the theoretical value of $1/2$. Nevertheless, because $\delta_u/\delta_\theta$ remains $O(1)$ throughout the entire range investigated, it is treated as a leading-order constant in the scaling analysis that follows. It is also plausible that the asymptotic value of $\delta_u/\delta_\theta$ depends on the Prandtl number, as commonly observed in classical Rayleigh--B\'enard convection. However, assessment of such a dependence requires simulations spanning a range of $\mathrm{Pr}$ and therefore lies beyond the scope of the present work. Within these limitations, Eq.~\eqref{eqn:epsilon_u_delta} supports the following leading-order scaling for the Reynolds number:
\begin{equation}
    \mathrm{Re} \sim \mathrm{Ra}_m^{1/2} \mathrm{Pr}^{-1},
    \label{eqn:scaling_Re}
\end{equation}
since $U = \mathrm{Re} \, \mathrm{Pr}$ by our choice of characteristic scales. Notably, substituting Eq.~\eqref{eqn:scaling_Re} into this relation shows that $U \sim \mathrm{Ra}_m^{1/2}$, meaning the large-scale velocity becomes independent of the Prandtl number.

We then proceed with the derivation of the scaling for the Nusselt number. Following~\cite{GROSSMANN_LOHSE_2000}, a further constraint can be derived from the energy equation~\eqref{eqn:energy}, which in two dimensions reads
\begin{equation}
    u_x\partial_x\theta + u_z\partial_z\theta=\partial_{zz}\theta,
    \label{eqn:energy_2D}
\end{equation}
where the streamwise diffusive contribution is neglected, since $\partial_{xx}\theta \ll \partial_{zz}\theta$. We note that the turbulent flow organizes into multicellular large-scale circulations (LSCs) with a pronounced vertical component, as shown in Fig.~\ref{fig:Fig2}(a). This structure is consistent with the presence of strong Kelvin forces in regions of large thermal gradients near the heated boundaries, which promote enhanced plume ejection and vertical thermal transport. Although localized plume ejection and recirculating motions generate non-negligible lateral convective heat transport, their contribution to the global heat transport is negligible on average compared with its wall-normal counterpart, i.e., $\langle u_x\partial_x\theta\rangle_A \ll \langle u_z\partial_z\theta\rangle_A$, reflecting the fact that positive and negative lateral transport events tend to compensate each other when averaged along the wall, whereas the wall-normal contribution retains a finite net heat flux across the thermal boundary layer. Consequently, using Eq.~\eqref{eqn:energy_2D} the leading-order transport balance can be expressed in terms of wall-normal advection and diffusion across the thermal-boundary-layer thickness, yielding ${U}/{\delta_\theta} \sim {1}/{\delta_\theta^2}$. Since $\mathrm{Nu}\sim\delta_\theta^{-1}$ and $U=\mathrm{Re}\,\mathrm{Pr}$ under the present nondimensionalization, it follows that
\begin{equation}
\mathrm{Nu} \sim \mathrm{Re}\,\mathrm{Pr}.
\label{eqn:energy_2D_vertical}
\end{equation}
Combining Eq.\eqref{eqn:scaling_Re} and Eq.\eqref{eqn:energy_2D_vertical}, we finally obtain
\begin{equation}
\mathrm{Nu} \sim \mathrm{Ra}_m^{1/2}.
\label{eqn:scaling_Nu}
\end{equation}

In the spirit of the Grossmann and Lohse theory~\cite{GROSSMANN_LOHSE_2000}, a complete picture of the transport requires partitioning the dissipation and scaling behaviors between the boundary layers and the bulk. To examine the scaling associated with the bulk contribution, we analyze the bulk alone; for the sake of convenience, we proceed with our considerations using the dimensional form of the governing equations. For $\mathrm{Ra}_m\gg1$, the Kelvin body force dominates over viscous effects in the bulk. Moreover, since the magnetic force must then be balanced by the remaining non-viscous terms, inertia and the pressure gradient are of the same order of magnitude, i.e., $\rho U^2/L\sim\Delta p/L$. Hence, a characteristic velocity scale for bulk plume motion can be estimated by balancing inertial advection and the Kelvin body force. The vertical Kelvin force scales as $f_{K,z}\sim \mu_v K\left(\theta-\theta_0 \right)\partial_zH$, while the component of inertia in the vertical direction is $\rho u_z \partial_z u_z \sim \rho u_z^2/L$. Balancing these two we obtain
\begin{equation}
    \rho \frac{u_z^2}{L} \sim \mu_v K\left(\theta-\theta_0 \right)\partial_zH.
        \label{eqn:_advective_balance}
\end{equation}
The induced vertical gradient of the magnetic field can be evaluated from the dimensional form of Eq.~\eqref{eqn:Maxwell} (see, e.g.,~\cite{suslov2008}), which provides $\partial_zH \sim K\partial_z\theta/(1+\chi)$. Thermal plumes ejected from the boundaries traverse the core along quasi-vertical trajectories while experiencing thermal variations on the order of the applied temperature gradient; hence, we estimate $\partial_z\theta \sim \beta$. The same structures carry temperature anomalies of order $\beta L$, since for sufficiently large P\'eclet numbers diffusive losses along their trajectories remain limited. Substituting these relations into Eq.~\eqref{eqn:_advective_balance} yields the characteristic vertical velocity associated with the motion of the thermal plumes:
\begin{equation}
    u_z \sim \left[\frac{\mu_v K^2\beta^2L^2}{\rho(1+\chi)} \right]^{1/2}.
        \label{eqn:u_zeta}
\end{equation}
From the definition of the Reynolds number $\mathrm{Re}=u_zL/\nu$, using expression~\eqref{eqn:u_zeta}, it follows that
\begin{equation}
 \mathrm{Re}\sim\mathrm{Ra}_m^{1/2}\mathrm{Pr}^{-1/2}.   
\end{equation}
Additionally, the predominantly advective transport of these thermal anomalies across the bulk suggests that the convective heat flux can be estimated as
\begin{equation}
q_{\mathrm{conv}}\sim \rho c_p u_z \beta L.
\label{eqn:q_zeta}
\end{equation}
Since $\mathrm{Nu}=q_\mathrm{conv}/q_\mathrm{cond}$, where $q_\mathrm{cond}\sim\rho c_p\kappa\beta$ is the conductive heat transport, Eq.~\eqref{eqn:q_zeta} implies $\mathrm{Nu}\sim u_zL/\kappa=\mathrm{Re}\mathrm{Pr}$, i.e.,
\begin{equation}
\mathrm{Nu}\sim\mathrm{Ra}_m^{1/2}\mathrm{Pr}^{1/2}.
\end{equation}

Because both the boundary-layer (Eq.~\eqref{eqn:scaling_Nu}) and the bulk contributions above exhibit a $\mathrm{Ra}_m^{1/2}$ dependence, any global transport law obtained from their combination preserves this asymptotic exponent. However, since the present DNS campaign is restricted to a single Prandtl number, we do not attempt to infer a global $\mathrm{Pr}$-scaling law. While the bulk analysis suggests a non-trivial Prandtl-number dependence in the asymptotic regime, its assessment requires dedicated simulations spanning a broader range of $\mathrm{Pr}$ and is therefore deferred to future work. Accordingly, at fixed Prandtl number,
\begin{equation}
    \mathrm{Nu}\sim\mathrm{Ra}_m^{1/2}, \quad \mathrm{Re}\sim\mathrm{Ra}_m^{1/2}.
    \label{eqn:scalings_Nu_Re}
\end{equation}
Remarkably, aside from the dependence on $\mathrm{Pr}$, Eqs.~\eqref{eqn:scalings_Nu_Re} exhibit the same functional dependence on the driving parameter as the asymptotic Kraichnan ultimate regime of classical thermal convection~\cite{kraichnan1962}. We emphasize, however, that this analogy is purely formal, as the underlying driving mechanism differs fundamentally from buoyancy-driven Rayleigh--B\'enard convection. In classical RB convection, this regime is expected only when boundary layers become fully turbulent at extreme Rayleigh numbers, as postulated by the original theory of Kraichnan and evidenced by experiments (see, e.g.,~\cite{Chavanne1997,Chavanne2001,He2012}) and recent two-dimensional DNS~\cite{Zhu2018}. In the present magnetic configuration, by contrast, half-power scaling laws emerge immediately upon transition to turbulence, despite the boundary layers remaining laminar. This mechanism is primarily governed by magnetic forcing, which enhances the ejection of thermal plumes from the thermal boundary layer and promotes their advection across the bulk. Since the thermal and viscous boundary-layer thicknesses are comparable (and at the lowest $\mathrm{Ra}_m$ the viscous layer is even thinner than the thermal one), newly generated plumes are immediately exposed to strong advective transport. This facilitates their extraction from the boundary region and their subsequent transport toward the opposite boundary, as shown in Fig.~\ref{fig:Fig2}(b). This process establishes a convective thermal ``short-cut'' between the heated boundaries, since plumes traverse the domain with limited thermal diffusion, thereby enhancing heat transport. 

Figure~\ref{fig:Fig3}(a) shows the Nusselt number, $\mathrm{Nu}$, as a function of the magnetic Rayleigh number, $\mathrm{Ra}_m$ (the raw $\mathrm{Nu}$ values together with their 95\% confidence intervals are reported in the Supplemental Material). Symbols denote the DNS data, while the dashed line represents the unconstrained power-law best fit, which yields an exponent slightly larger than the theoretical prediction of $1/2$. The compensated representation $\mathrm{Nu}/\mathrm{Ra}_m^{1/2}$ shown in the inset of Fig.~\ref{fig:Fig3}(a) nevertheless suggests a tendency toward a half-power scaling regime, although the limited span of data does not allow definitive conclusions. Given the relatively restricted range of $\mathrm{Ra}_m$ explored, these trends should be interpreted with due caution, and further investigations at higher $\mathrm{Ra}_m$ will be required to conclusively establish the asymptotic scaling behavior. A similar comparison for the Reynolds number is shown in Fig.~\ref{fig:Fig3}(b). The unconstrained power-law fit again yields an exponent larger than the theoretical value of $1/2$, while the compensated representation $\mathrm{Re}/\mathrm{Ra}_m^{1/2}$ displayed in the inset suggests a gradual approach toward a plateau. Remarkably, despite the relatively low values of $\mathrm{Re}$, which are of order $O(10^0)$--$O(10^1)$, the large Prandtl number yields a P\'eclet number $\mathrm{Pe}=\mathrm{Re}\,\mathrm{Pr}$ of order $O(10)$--$O(10^2)$, indicating that thermal transport is already predominantly advective in the bulk. It is worth noting that both fitted exponents exceed the theoretical prediction over the range investigated. We speculate that this discrepancy may be attributed, on the one hand, to the fact that the flow regime is still evolving and may not yet have reached its asymptotic state (cf. the previous discussions about the evolution of the spectra and the boundary layer thickness), and, on the other hand, to the two-dimensional nature of the simulations and the presence of lateral confinement. In two-dimensional flows, the absence of vortex stretching fundamentally alters the transfer of kinetic energy across scales, promoting the formation of long-lived coherent structures compared to three-dimensional flows. In addition, lateral solid boundaries favor the emergence of large-scale corner rolls, which enhance flow organization and may further contribute to the observed departure from the asymptotic scaling behavior. We stress that these interpretations remain conjectural, and a systematic assessment based on fully three-dimensional simulations will be required to clarify their role.

In conclusion, our study demonstrates that magnetic forcing can trigger the early emergence of half-power transport scaling while both thermal and viscous boundary layers remain laminar. These findings suggest that enhanced convective transport need not be intrinsically linked to turbulent boundary layers, opening new perspectives for the study of extreme transport phenomena in magnetically active fluids.

The author wishes to thank Professor Marcello Lappa for useful discussions and a careful revision of the manuscript.

\nocite{*}
\bibliography{aipsamp}

\end{document}